\documentclass[11pt,sort&compress]{elsarticle}   	
\usepackage{amssymb,color}

\usepackage{lineno,hyperref,geometry,amsmath,appendix}
\modulolinenumbers[5]
\usepackage{bbold}
\newcommand{\gdualn}[1]{\overset{\:{}^{{}^{\boldsymbol{\neg}}}}{\smash[t]{#1}}} 
\def\0{\mbox{\boldmath$\displaystyle\mathbb{O}$}}

\def\kb{\mbox{\boldmath$\displaystyle\boldsymbol{\kappa}$}}
\def\bz{\mbox{\boldmath$\displaystyle\boldsymbol{\zeta}$}}

\def\I{\openone}

\def\p{\mbox{\boldmath$\displaystyle\boldsymbol{p}$}}
\def\0{\mbox{\boldmath$\displaystyle\boldsymbol{0}$}}

\def\openone{\mathbb I}

\journal{European Physical Journal -- Special Topics}

\begin{document}

\begin{frontmatter}
\title{\textsc{\textcolor{black}{Elko and Mass Dimension One Fermions: Editorial}}}

\author[mymainaddress]{Dharam Vir Ahluwalia\corref{mycorrespondingauthor}}
\cortext[mycorrespondingauthor]{Corresponding author}
\ead{dharam.v.ahluwalia@gmail.com}

\address[mymainaddress]{Mountain Physics Camp, Center for the Studies of the Glass Bead Game\\  Bir, Himachal Pradesh, 176077
India}


\begin{abstract}
\textcolor{black}{This editorial presents the quantum field theoretic ambiance in which Elko and mass dimension one fermions come to exist. This may serve not only to introduce Elko, and the associated quantum field, but it may also open a door to a new class of spin one half fermions and bosons. These new particles, as it will be argued in this volume, are first-principle candidates for dark matter. }
\end{abstract}

\end{frontmatter}
\vspace{21pt}

Our reader certainly needs no reminder that astrophysical and cosmological observations reveal that the Standard Model of high energy physics (SM-hep)  accounts for only 
 $5\%$ of the universe.  The remaining 
 $95\%$ is dark to us, except gravitationally. It is bifurcated into dark matter, $25\%$,  and dark energy, $70\%$. 
 The \emph{kinematic structure} of the SM-hep fermions allows introduction of certain \emph{local gauge symm\-etries}. These account for the `luminosity' of our universe, of stability of nuclei and atoms, and radioactivity. All chemical and nuclear reactions arise from the fields associated with the local gauge symmetries -- with gravity being an important source of astrophysical and cosmological structures, and primary initiator of astrophysical nuclear reactions. The essential elements of the kinematic structure are  the spin-$1/2$ quantum fields of Dirac, and the Dirac operator $\left(i \gamma_\mu \partial^\mu - m \I\right)$ -- and, the definition of spinorial dual and adjoint. In momentum space, $m^{-1} \gamma_\mu p^\mu$ serves as the parity operator. Modulo certain phases, its eigenfunctions serve as the expansion coefficients of the quantum field of Dirac. 
 
Taking a hint from 
the standard model of high energy physics (SM-hep) we together may ask:
 is the kinematic structure of the dark matter based on the quantum-field theoretic extension of  Dirac's 1928 formalism in such a manner that it  does not support the gauge
symmetries of the SM-hep. Is there only one dark-matter sector, or there are several? 
The central thesis of this issue of EPJ-ST is to answer the asked question in the affirmative, and to open up a logical possibility that there may not be one but several dark-matter sectors. A result, on the way,  would suggest  how to tame  zero point 
energies of all the cosmic matter fields. 
In undertaking our task we prefer to be conservative: whatever dark matter is it must be one representation or the other of the Lorentz algebra, and the well known discrete symmetries. This is our working conjecture.

The question we ask would seem to be blocked by a no-go theorem inherent in the 1964 work of Weinberg~\cite{PhysRev.133.B1318,Weinberg:1995mt}. The block is evaded by the realisation that the duals of the spinors in the $(1/2,0)\oplus(0,1/2)$ representation space are not unique, nor are the adjoints~\cite{Ahluwalia:2004ab,Ahluwalia:2009rh,Ahluwalia:2019etz,Cavalcanti_2020}. And, arguments akin to those of Carl Bender further help evade the no-go theorem~\cite{PhysRevLett.80.5243,Ahluwalia:2016jwz} . 
\vspace{11pt}

The story of the SM-hep matter 
 begins with the 1928  paper of Dirac~\cite{Dirac:1928hu}. At its core, it presents a non-trivial square root of the dispersion relation
$p_\mu p^\mu = m^2$. The square root of the left-hand side that Dirac obtained was $\gamma_\mu p^\mu$. The square root of the right-hand side was implicitly  assumed to be $\pm\, m\, \I$, where $\I$ is an identity matrix in the four dimensional $(1/2,0)\oplus(0,1/2)$ representation space. Combined, these yield the momentum-space wave equation 
$\left(\gamma_\mu p^\mu \pm m\,\I\right)\psi(\p)=0$.

A first hint that the kinematic structure suggested by Dirac may not be unique arises from two observations: (a) casting Dirac's 1928 work 
in the manner just done,
 immediately tells us that non-trivial square roots of $4\times 4$ identity matrix $\I$ may lead to fundamentally new kinematic structures;
(b)  in 2014, Speran\c{c}a observed that $m^{-1}\gamma_\mu p^\mu$ serves as the parity operator, with the consequence that 
 $\psi(\p)$ emerge as eigenspinors of the parity operator~\cite{Speranca:2013hqa}.\footnote{The parity operator $m^{-1}\gamma_\mu p^\mu$  is derived without reference to Dirac equation, see~\cite[Section 5.3]{Ahluwalia:2019etz} and~\cite{Speranca:2013hqa}.}
A first example of a new kinematic structure arising from observation (a) has just been published~\cite{Ahluwalia:2020PRSA}. 
Observation (b) led to the theoretical  discovery of Elko -- eigenspinors of the charge conjugation operator --   and mass dimension one fermions. 

To clear a path for our reader, we note that there are two extremes in the manner in which Dirac's quantum field is introduced to the un-initiated.
 In some of these works `derivation' of Dirac equation contains a serious mistake~\cite[page 167]{Hladik:1999tt} and~\cite[page 44, in the first edition]{Ryder:1985wq}, and the pairing between the  the annihilation operators and the antiparticle spinors is incompatible with that required by parity covariance, rotational symmetry, and cluster decomposition principle. The other extreme is Weinberg's monograph on the subject. It is flawless but it requires a serious investment of time. 
 
 In the notation of~\cite{Ahluwalia:2019etz}, the mistake in the derivation of Dirac equation lies in the fact that for the particle spinors one indeed has $\phi_R(\p=\0,\pm \sigma)= + \phi_L(\p=\0,\pm \sigma)$ but for the antiparticle spinors one must have $\phi_R(\p=\0,\pm \sigma)= - \phi_L(\p=\0,\pm \sigma)$~\cite{Gaioli:1998ra,Ahluwalia:1998dv}. The correct pairing, upto a common normalisation factor,  of spinors and phases for the particle and antiparticle spinors must be as follows~\cite[page 224]{Weinberg:1995mt}:

\begin{align}
u(\p,+1/2) =+\, \kb \frac{1}{\sqrt{2}} 
\left(
\begin{array}{c}
1\\
0\\
1\\
0
\end{array}
\right), \quad \mbox{paired with}~ a(\p,+1/2) \\
u(\p,-1/2) =+\, \kb \frac{1}{\sqrt{2}} 
\left(
\begin{array}{c}
0\\
1\\
0\\
1
\end{array}
\right), \quad \mbox{paired with}~ a(\p,-1/2)
\end{align}
and
\begin{align}
v(\p,+1/2) =+\,\kb \frac{1}{\sqrt{2}} 
\left(
\begin{array}{l}
0\\
1\\
0\\
-1
\end{array}\label{eq:va}
\right), \quad   \mbox{paired with}~ b^\dagger(\p,+1/2)\\
v(\p,-1/2) = - \,\kb \frac{1}{\sqrt{2}} 
\left(
\begin{array}{c}
1\\
0\\
-1\\
0
\end{array}
\right),\quad  \mbox{paired with}~ b^\dagger(\p,-1/2)
\label{eq:vb}
\end{align}
What goes wrong is that in (\ref{eq:va}) and  (\ref{eq:vb}), $v(\p,+1/2)$ and $v(\p,-1/2)$ are interchanged (see, for example~\cite[page 113]{Folland:2008zz} and a fair fraction of textbooks on quantum field theory); without altering the right-hand sides.\footnote{I am grateful to Gerald Folland for a private communication, and his permission to point out the error.} In addition, both the multiplicative  phase factors that appear before $\kb$  are made $+1$.
The boost operator $\kb$ in the above equations is
\begin{equation}
\kb
 = \sqrt{\frac{E + m }{2 m}}
						\left[
						\begin{array}{cc}
						\I + \frac{\boldsymbol{\sigma}\cdot\mathbf{p}}{E +m} & \0 \\
						\0 & \I - \frac{\boldsymbol{\sigma}\cdot\mathbf{p}}{E +m} 
						\end{array}
						\right]  \label{eq:sb-again}
\end{equation}
Thus the particle and antiparticle spinors satisfy Dirac equation. Each of them has a freedom of a multiplicative phase factor. These phase factors can vary from one spinor to another but when we use these spinors as expansion coefficients of a quantum field, these become relative -- with additional constraint entering on Majorana-isation of the field. The cluster decomposition principle, Lorentz and Parity covariance fix these phases to be as above. These observations, in part, led to the removal~\cite{Ahluwalia:2016rwl,Ahluwalia:2019etz} of the non-locality present in the early works on mass dimension one fermions (with Elko as expansion coefficients)~\cite{Ahluwalia:2004sz,Ahluwalia:2004ab}.

This discussion sets the ambiance for the kinematic structure of known matter fields  -- that is, the field necessary for all the SM-hep matter fields. We assume that dark-matter sector(s), like its cousin in the SM-hep, is (are) also described by spin one half particles. We say particles because spin-$1/2$ supports not only an entirely new class of mass dimension one fermions but also bosons! Since it was too late to include a detailed review article on the subject, 
the reader is referred to two publications in press~\cite{Ahluwalia:2020miz,Ahluwalia:2020jkw}

To bring this editorial to a close, we define a generalised quantum field of the form
\begin{equation}
\psi(x) \stackrel{\mathrm{def}}{=} 
\int
\frac{d^3p}{(2\pi)^{3/2}}  \sum_\sigma\left[ 
u(\p,\sigma)e^{-i p\cdot x} a(\p,\sigma)
+
v(\p,\sigma)e^{i p\cdot x} b^\ddagger(\p,\sigma)
\right]
\end{equation}
where
\begin{align}
u(\p,\sigma)~ \mbox{and}~ v(\p,\sigma) = & ~ \mbox{a orthonormal set of expansion coefficients} \nonumber \\ & \mbox{~in the finite dimensional representtaion space}\nonumber \\ & \mbox{~of the Lorentz algebra}\nonumber
\end{align}
and $\ddagger$ is an adjoint (which for the purposes of this editorial shall be taken as the usual hermitian adjoint $\dagger$). The $a(\p,\sigma)$ and $a^\dagger(\p,\sigma)$, and similarly $b(\p,\sigma)$ and $b^\dagger(\p,\sigma)$, satisfy the usual bosonic or fermionic commutators or anticommutators.
The dual of the expansion coefficients, jointly called $\xi(\p,\sigma)$, is defined as
\begin{equation}
\gdualn{\xi}(\p,\sigma) = \left[\Xi(\p) \xi(\p,\sigma)\right]^\ddagger \eta \end{equation}
where $\Xi(\p)$ is an invertible map  with  $\Xi(\p)^2 = \I$, and $\eta$ is determined by the constraints
\begin{align}
\left[\zeta_i, \eta \right] =0,\quad i=x,y,z;\qquad
\left\{\kappa_i, \eta \right\} =0\quad i=x,y,z
\end{align}
The $\bz$ and $\kb$ are the generators of rotations and boosts in the representation space in which the $\xi(\p,\sigma)$ reside. For spin one half, setting $\Xi(\p) = \I, \forall\, \p$, one recovers the usual Dirac dual of spinors, and the standard adjoint. 

Original mass dimension one fermions arose from taking the 
$u(\p,\sigma)$ and $v(\p,\sigma)$ as eigenspinors of the charge conjugation (Elko). It was discovered that under the Dirac dual these massive spinors had a norm that identically vanished. This observation motivated the introduction of a new dual first introduced in~\cite{Ahluwalia:2003jt}, and later taken to its final form in~\cite{Ahluwalia:2016rwl,Ahluwalia:2019etz}.\footnote{That said, the subject is still under active investigation. We refer the reader to~\cite{Cavalcanti_2020} and references therein for a full flavour of current research activity.} In a series of papers it was argued that the newly discovered fermions carried a natural darkness with respect to the SM-hep fields. In addition they had a natural dimension four quartic self interaction. First arguments to this effect can be traced back to~\cite{Ahluwalia:2004sz,Ahluwalia:2004ab}.

Editor's monograph provides an elementary and pedagogic introduction to Elko and mass dimension one fermions~\cite{Ahluwalia:2019etz}. Equipped with that knowledge Cheng-Yang Lee's opening review to this volume provides a slightly different perspective and extends the formalism to higher spins~\cite{Lee:2019fni}. 
One loop calculations are presented in two papers. First of these is the paper by Gustavo Pazzini de Brito and colleagues~\cite{deBrito:2019hih}, the second is the contribution by Alekha Nayak and Tripurari Srivastava~\cite{Nayak:2020ntm}. Elko have drawn significant attention of studies on branes. The review by  Xiang-Nan Zhou and Yu-Xiao Liu is devoted to this emerging field of research~\cite{Zhou:2020ucc}. Following this
Saulo Pereira et al. review the cosmology based on the new spinors~\cite{Pereira:2020ogo}. It is intended to serve as a primer on Elko cosmology. It also gives references to earlier pioneering studies on the subject. Following this are two papers, one by Luca Fabbri who looks at Elko in polar form~\cite{Fabbri:2019vut}. The second, by Marco Dias et al. concludes this volume by presenting a 
tutorial approach to phenomenological analysis 
 in the context of mass dimension one fermions~\cite{Duarte:2020svn}.

\vspace{7pt}

This editorial only sets the stage for the invited authors to sow seeds of their contributions. The stage is now theirs, and I hope we have succeeded in making the stage well-lighted for the readers, our audience.

\vspace{21pt}

\noindent
\textbf{Acknowledgements.} 
It is now time to thank Balasubramanian Ananthanarayan who first envisioned a need for this volume and invited me to consider composing an EPJ-ST volume on Elko and Mass dimension one fermions. I thank him, and also  Sandrine Karpe of the editorial and production offices of EPJ-ST for attending to various formalities and needs. Thanks are also extended to all the referees, and to all the authors for making this project take this form.  And thank you Sweta Sarmah for all your good contributions to this project. Again, thank you all: cheers!
\vspace{7pt}

\noindent
\textbf{Funding.} The research presented here is entirely supported by the personal funds of the author.

\vspace{21pt}

\noindent

\end{document}